\documentclass[aps,prb,superscriptaddress,amsmath,amssymb,floatfix,twocolumn,showpacs,amsfonts,floatfix]{revtex4-1}
\usepackage{txfonts}
\usepackage{textcomp}
\usepackage{graphicx}
\usepackage{subfigure}
\usepackage{color}
\usepackage[colorlinks=true,citecolor=blue,urlcolor=blue,linkcolor=blue]{hyperref}
\usepackage{braket}
\usepackage{overpic}

\def\kk{\mathbf{k}}
\def\qq{\mathbf{q}}
\def\bs{\mathbf{S}}

\def\bdelta{\mathbf{\delta}}

\def\half{\frac{1}{2}}

\allowdisplaybreaks

\begin{document}

\title{Spectrum splitting of bimagnon excitations in a spatially frustrated Heisenberg antiferromagnet revealed by resonant inelastic x-ray scattering}

\author{Cheng Luo}
\affiliation{State Key Laboratory of Optoelectronic Materials and Technologies,\\School of Physics and Engineering, Sun Yat-sen University, Guangzhou 510275, China}

\author{Trinanjan Datta}
\email[Corresponding author:]{tdatta@gru.edu}
\affiliation{Department of Chemistry and Physics, Georgia Regents University, 1120 15th Street, Augusta, GA, 30912}
\affiliation{State Key Laboratory of Optoelectronic Materials and Technologies,\\School of Physics and Engineering, Sun Yat-sen University, Guangzhou 510275, China}

\author{Dao-Xin Yao}
\email[Corresponding author:]{yaodaox@mail.sysu.edu.cn}
\affiliation{State Key Laboratory of Optoelectronic Materials and Technologies,\\School of Physics and Engineering, Sun Yat-Sen University, Guangzhou 510275, China}

\date{\today}

\begin{abstract}
We perform a comprehensive analysis of the bimagnon resonant inelastic x-ray scattering (RIXS) intensity spectra of the
spatially frustrated $J_x-J_y-J_2$ Heisenberg model on a square lattice in both the antiferromagnetic and the collinear antiferromagnetic phase.
We study the model for strong frustration and significant spatial anisotropy to highlight the key signatures of RIXS spectrum splitting which
may be experimentally discernible. Based on an interacting spin wave theory study within the ladder approximation Bethe-Salpeter scheme, we find the appearance of a robust two-peak
structure over a wide range of the transferred momenta in both magnetically ordered phases.
The unfrustrated model has a single-peak structure with a two-peak splitting originating due to spatial anisotropy and frustrated interactions.
Our predicted two-peak structure from both magnetically ordered regime can be realized in iron pnictides.
\begin{description}
\item[PACS number(s)] 78.70.Ck, 75.25.-j, 75.10.Jm
\end{description}
\end{abstract}

\maketitle

\section{Introduction}
Resonant inelastic x-ray scattering (RIXS) has recently been established as a powerful spectroscopic technique to study elementary
excitations in strongly correlated electron materials.~\cite{RevModPhys.83.705}
The energy of the incoming X-ray photon is resonantly tuned to match an element absorption edge,
thus generating a large enhancement of the scattered intensity. As incident radiation loses its energy and momentum to excitations
inherent to the material, direct information on the dispersions of spin,
~\cite{PhysRevLett.100.097001,PhysRevLett.102.167401,PhysRevLett.103.047401,PhysRevLett.104.077002,PhysRevLett.105.157006,PhysRevLett.107.107402,
NatPhys.7.725,PhysRevLett.108.177003,PhysRevB.85.214527,NatMater.11.850,NatMater.12.1019}
orbital,~\cite{PhysRevLett.101.106406,PhysRevLett.103.107205,Nature.485.82,PhysRevLett.109.117401} lattice,~\cite{JPCM.22.485601,PhysRevLett.110.265502}
and other degrees of freedom~\cite{PhysRevLett.110.087403,PhysRevLett.110.117005} can be obtained.

Magnetic correlations give rise to both single and double spin-flip excitations, which are equivalent to single and bimagnon
excitations in long-range ordered Heisenberg antiferromagnet. Measuring single spin-flip excitation has traditionlly been the domain of inelastic neutron
scattering,~\cite{PhysRevLett.86.5377,nature02576}
however, {\it direct} RIXS experiment at $2p\leftrightarrow3d$ edges of Cu via strong spin-orbital coupling in the core state has started to challenge
this monopoly $-$ with the additional advantage of requiring only small sample sizes and probing the entire Brillouin zone (BZ).
~\cite{PhysRevLett.103.117003,PhysRevLett.105.167404}
In the {\it indirect} process at Cu K edges ($1s\leftrightarrow4p$), RIXS can create double spin-flip excitations,
where the single spin-flip scattering is forbidden since the total spin of the valence electrons is conserved.
The microscopic mechanism underlying bimagnon excitations
involves a local modification of the superexchange interaction mediated via the core hole,  thus leading to the RIXS
spectra expressed as a momentum-dependent four-spin correlation function.~\cite{EPL.80.47003,PhysRevB.75.214414,PhysRevB.77.134428,NJP.11.113038}
This makes RIXS complementary to optical Raman scattering, which also measures the bimagnon excitations,
but restricted to zero momentum transfer.~\cite{RevModPhys.79.175}
Despite the success of disentangling both single and bimagnon excitations in a variety of antiferromagnetic (AF) ordered cuprates,
there are several two-dimensional (2D) materials of interest that are yet to be studied with RIXS.
Furthermore, superconductivity in the iron pnictides exists in close vicinity of the collinear antiferromagnetic (CAF) order~\cite{NatPhys.5.555}
where single magnon excitations have been proposed and observed with {\it direct} RIXS at Fe L edge very recently.~\cite{PhysRevB.84.020511,NatCommun.4.1470}
Both the frustrated $J_1-J_2$ model~\cite{PhysRevB.78.052507,PhysRevB.79.092416,PhysRevB.84.155108}
and the spatially anisotropic $J_x-J_y-J_2$ model~\cite{PhysRevB.81.024505,PhysRevB.81.165101}
can support the $(\pi,\pi)$-AF and the $(\pi,0)$-CAF phase,
with complex vanadium oxide compounds serving as additional excellent material realizations.~\cite{PhysRevB.79.214417}

In this paper, we investigate the key signatures of strong spatial anisotropy and magnetic frustration in the {\it indirect}
RIXS spectra of the $J_x-J_y-J_2$ model in both the AF and the CAF ordered phase.
We compute the bimagnon RIXS intensity including up to first-order $1/S$ spin wave expansion
correction with the ladder approximation Bethe-Salpeter equation.
We find the appearance of a robust two-peak structure over a wide range of transfered momenta $\qq$ inside the BZ.
In the AF phase, the presence of both spatial anisotropy and next-neighbor frustration is the key to spectrum splitting of the
RIXS intensity spectra. The spectrum splitting originates in the pure spatially anisotropic $J_x-J_y$ model. Introducing a small frustration
on the basis of the anisotropic model can lead to significant peak splitting (see Figs.~\ref{fig:afcaffall} and ~\ref{fig:afcafky} ).
We also find that the reduced quantum fluctuation from large spin values lead to the disappearance of this structure.
For the CAF phase, the presence of strong magnetic frustration induces the spectrum splitting. In the nonfrustrated model ($J_y\lesssim0$),
we find a single peak. For weak frustrated interaction the two-peak structure develops which is not stable and can be destroyed by increasing spin values,
while for strong frustration the two-peak structure is still robust even in the non-interacting ($S\rightarrow\infty$) regime.
Furthermore, similar to Raman we find a spectral downshift caused by increasing the frustrated interaction
($J_2$ for the AF phase while $J_y$ for the CAF phase).~\cite{PhysRevLett.106.067002}

This article is organized as follows. In Sec.~\ref{Sec:Model} we introduce our model Hamiltonian with $1/S$ spin-wave expansion correction.
In Sec.~\ref{Sec:rixs} we introduce the indirect RIXS process with explicit expressions for the scattering operator in both the AF and the CAF phase.
In Sec.~\ref{Sec:intensity} we report our results of the RIXS intensity spectrum and describe the relation between our theoretical spectrum
and the proposed experimental features in the CAF ordered phase of iron pnictides.
In Sec.~\ref{Sec:ins} we include a comparison, within our theoretical approach, of the bimagnon excitations probed by Raman and
Inelastic Neutron Scattering (INS) spectroscopic techniques to the RIXS method.
Finally in Sec.~\ref{Sec:Conclu} we present our discussions and concluding remarks.

\section{Model Hamiltonian}\label{Sec:Model}

\begin{figure}
\centering\includegraphics[scale=0.6]{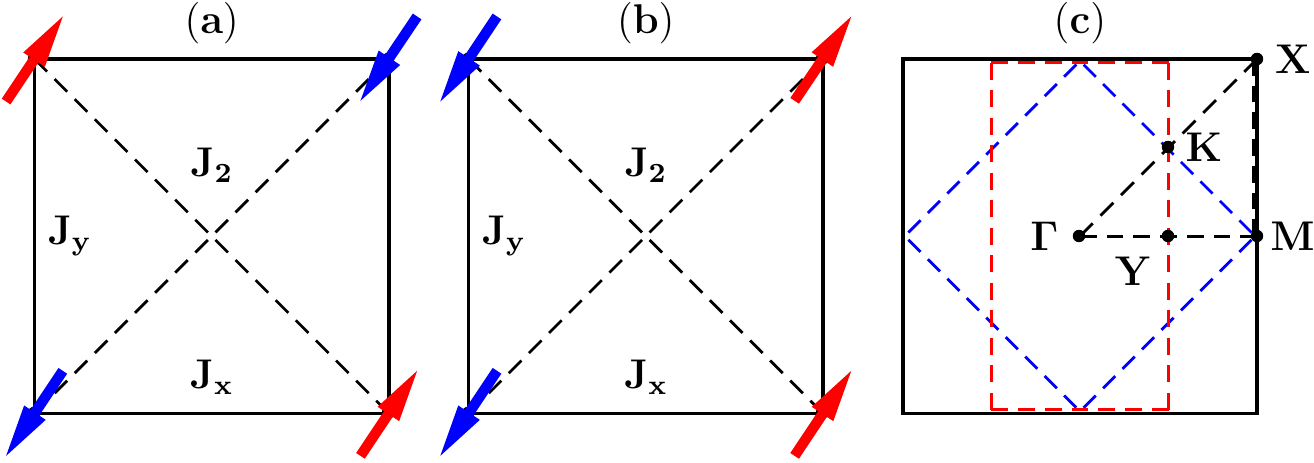}
\caption{(Color online)
Schematic representation of $J_x-J_y-J_2$ model. (a) AF ordered phase, (b) CAF ordered phase, (c) Brillouin Zone (BZ).
Dashed lines in (c) represent the magnetic BZ boundary for AF phase (blue) and CAF phase (red), respectively. Coordinates of the points
in the BZ are $\Gamma(0,0)$, $K(\frac{\pi}{2},\frac{\pi}{2})$, $X(\pi,\pi)$, $M(\pi,0)$, $Y(\frac{\pi}{2},0)$.
}
\label{fig:Model}
\end{figure}

The $J_x-J_y-J_2$ Hamiltonian is given by
\begin{equation}
  \mathcal{H}=\sum_{ij}J_{ij}\bs_i\cdot\bs_j,
\end{equation}
where $\bs_i$ is the spin on site i, $J_{ij}$ is $J_x$ along the nearest-neighbor (NN) x (row) direction, $J_y$ is along the NN y (column) direction,
and $J_2$ is the next-NN interaction along the diagonals in the xy plane. The dimensionless ratios $\zeta=J_y/J_x$ and $\eta=J_2/J_x$ denote the
relative interaction strength. At zero temperature the classical ground states for the spatially frustrated model are the Ne\'{e}l
ordered AF state and the CAF state, as shown in Fig.~\ref{fig:Model}.

We utilize the two sub-lattice Holstein-Primakoff transformation to bosonize the spin Hamiltonian where $a$ ($b$) bosons represent the
up (down) A (B) sublattices.~\cite{PhysRevB.46.10763,JPSJ.62.4449,PhysRevB.72.014403,PhysRevB.82.144407}
This is followed by a Fourier transformation to recast $\mathcal{H}$ in terms of the $a_\kk$ and $b_\kk$ bosons,
where $\kk$ is the wave-vector in the BZ. The original Hamiltonian, $\mathcal{H}$, can be written in momentum space as a sum of classical energy,
a quadratic term, and a quartic interaction term. We then diagonalize the quadratic part $H_0$ by transforming the operators
$a_\kk$ and $b_\kk$ to magnon operators $\alpha_\kk$ and $\beta_\kk$ using the Bogoliubov transformations $a_\kk^\dag=u_\kk\alpha_\kk^\dag+v_\kk\beta_\kk, b_\kk=v_\kk\alpha_\kk^\dag+u_\kk\beta_\kk$ where the coefficients
$u_{\kk}$ and $v_{\kk}$ are defined as
$u_{\kk}=\Big[\frac {1+\epsilon_{\kk}}{2\epsilon_{\kk}}\Big]^{1/2},
v_{\kk}=-\mathrm{sgn}(\gamma_{\kk})\Big[\frac{1-\epsilon_{\kk}}{2\epsilon_{\kk}} \Big]^{1/2}$
with
$\epsilon_{\kk}=(1-\gamma_{\kk}^2)^{1/2},\gamma_{\kk}=\frac{\gamma_{1\kk}}{\kappa_{\kk}},
\gamma_{1{\kk}}=\frac{\cos (k_x)+\zeta \cos (k_y)}{(1+\zeta)},
\gamma_{2{\kk}}=\cos (k_x)\cos(k_y),\kappa_{\kk}=1- \frac {2\eta}{1+\zeta} (1-\gamma_{2\kk})$.
After these transformations we obtain the \emph{renormalized} dispersion in the AF phase as $\omega_\kk=2J_xS(1+\zeta)(\kappa_{\kk}\epsilon_{\kk}+\frac{A_\kk}{2S})$
where $S$ is the value of spin. The $1/S$ order term coming from the one-loop correction of the quartic interactions~\cite{PhysRev.117.117,PhysRevB.82.144407} reads as
\begin{align}
A_{\kk}&= A_1 \frac {1}{\kappa_{\kk}\epsilon_{\kk}}\left[\kappa_{\kk}
-\gamma_{1{\kk}}^2\right] + A_2 \frac {1}{\epsilon_{\kk}}\left[1-\gamma_{2\kk}\right],
\end{align}
with
\begin{eqnarray}
A_1 &=& \frac{2}{N} \sum_{\kk} \frac 1{\epsilon_{\kk}}
\left[\frac {\gamma_{1{\kk}}^2}{\kappa_{\kk}}+\epsilon_{\kk}-1\right],\\
A_2 &=& \left(\frac {2\eta}{1+\zeta} \right)\frac{2}{N} \sum_{\kk}
\frac 1{\epsilon_{\kk}}\left[1-\epsilon_{\kk}-\gamma_{2{\kk}}\right].
\end{eqnarray}

For the CAF phase a similar calculation leads to the structure factors $\gamma_{1{\kk}}^\prime,\gamma_{2{\kk}}^\prime$
along with other quantities required for the calculations are defined as
$\epsilon_{\kk}^\prime=(1-\gamma_{\kk}^{\prime2})^{1/2},
\gamma_{\kk}^\prime=\frac{\gamma_{1{\kk}}^\prime}{\kappa_{\kk}^\prime},
\gamma_{1{\kk}}^\prime=\frac{\cos(k_x)[1+2\eta \cos (k_y)]}{(1+2\eta)},
\gamma_{2{\kk}}^\prime=\cos(k_y),
\kappa_{\kk}^\prime=1- \frac {\zeta}{1+2\eta} (1-\gamma_{2{\kk}}^\prime)$.
For the collinear phase the dispersion takes the form $\omega_\kk^\prime=2J_xS(1+2\eta)(\kappa^\prime_{\kk}\epsilon^\prime_{\kk}+\frac{A_\kk^\prime}{2S})$.
The coefficients for the $1/S$ correction~\cite{PhysRev.117.117,PhysRevB.82.144407} that appear in the dispersion are
\begin{align}
A_{\kk}^\prime&= A_1^\prime \frac {1}{\kappa_{\kk}^\prime\epsilon_{\kk}^\prime}\left[\kappa_{\kk}^\prime
-\gamma_{1{\kk}}^{\prime 2}\right] + A^\prime_2 \frac {1}{\epsilon^\prime_{\kk}}\left[1-\gamma^\prime_{2\kk}\right],
\end{align}
with
\begin{eqnarray}
A^\prime_1 &=& \frac{2}{N} \sum_{\kk} \frac 1{\epsilon^\prime_{\kk}}
\left[\frac {\gamma_{1{\kk}}^{\prime 2}}{\kappa^\prime_{\kk}}+\epsilon^\prime_{\kk}-1\right], \\
A^\prime_2 &=& \left(\frac {\zeta}{1+2\eta} \right)\frac{2}{N} \sum_{\kk}
\frac 1{\epsilon^\prime_{\kk}}\left[1-\epsilon^\prime_{\kk}-\gamma^\prime_{2{\kk}}\right].
\end{eqnarray}

The lowest $1/S$ order irreducible quartic interaction vertex with the fixed total momentum $\qq$ of a magnon pair,
which conserves the number of magnons in the scattering process, is of the form
\begin{align}
\mathcal{V}^{\alpha\beta}=\sum_{\kk\kk^\prime}\mathcal{V}_{\kk\kk^\prime}
\alpha^\dag_{\kk^\prime+\qq} \alpha_{\kk+\qq} \beta^\dag_{\kk^\prime} \beta_{\kk},
\end{align}
where we have recast the momentum-dependent factors, $\mathcal{V}_{\kk\kk^\prime}$, in the following separable form
\begin{equation}
\mathcal{V}_{\kk\kk^\prime}=\frac{2}{N}\sum_{\mathrm{m,n}=1}^{\mathrm{N_c}} v_{\mathrm{m}}(\kk) \Gamma_{\mathrm{m n}} v_{\mathrm{n}}(\kk^\prime),
\end{equation}
which creates $N_c=18$ channels of interaction in both the AF and the CAF phase
(see Appendix~\ref{Ladder Vertex} for explicit forms of the expressions for $v_\mathrm{m}, v_\mathrm{n}$, and $\Gamma_{\mathrm{mn}}$).

\section{Indirect RIXS process}\label{Sec:rixs}

\begin{figure}
\centering
\includegraphics[scale=0.55]{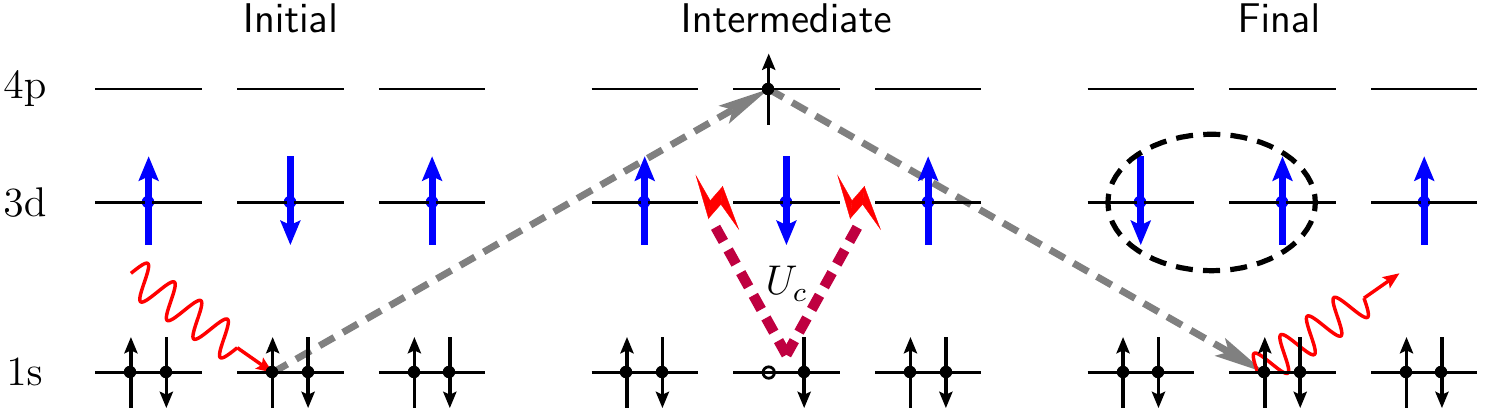}
\caption{(Color online) Illustration for a double spin-flip excitations in the {\it indirect} RIXS process at a transition metal K edge.}
\label{fig:1s4p}
\end{figure}

In the {\it indirect} RIXS process, e.g. transition metal K absorption edge, an inner 1$s$ electron is promoted to the 4$p$ band by absorbing a photon.
The presence of the localized core-hole potential $U_c$ in the intermediate state modifies the 3$d$ on-site Coulomb repulsion $U$; this perturbation
modifies the superexchange integral of the neighboring 3$d$ electrons (see Fig.~\ref{fig:1s4p}), yielding the two-magnon excitations.
We employ the ultrashort core-hole lifetime (UCL) expansion,~\cite{vandenBrink20052145,EPL.73.121,PhysRevB.75.115118} with the lowest order bimagnon
RIXS scattering operator~\cite{PhysRevLett.106.157205,PhysRevB.83.245133}
\begin{equation}\label{rixsop}
  \mathcal{O}_\qq=\sum_{ij}e^{i\qq\cdot \mathbf{r}_i}J_{ij}\bs_i\cdot\bs_j
\end{equation}
which can be expressed in terms of the bosonic quasiparticle form as
\begin{equation}
\mathcal{O}_\qq=\sum_{\kk}\mathcal{M}_{\kk,\qq}(\alpha_{\kk+\qq}^\dag\beta_{\kk}^\dag+\alpha_{\kk}\beta_{\kk+\qq}),
\end{equation}
with the following definition for the $\mathcal{M}_{\kk,\qq}$ factors in the AF phase
\begin{eqnarray}
\mathcal{M}^{\mathrm{AF}}_{\kk,\qq}&=&2J_x S(1+\zeta)\{[1+\gamma_{1\qq}+\frac{2\eta}{1+\zeta}\nonumber\\
&&\times(\gamma_{2\kk}+\gamma_{2\kk+\qq}-\gamma_{2\qq}-1)](u_{\kk+\qq}v_{\kk}+u_{\kk}v_{\kk+\qq})\nonumber\\
&&+(\gamma_{1\kk}+\gamma_{1\kk+\qq})(u_{\kk}u_{\kk+\qq}+v_{\kk}v_{\kk+\qq})\}.
\end{eqnarray}
The CAF phase RIXS operator expression can be obtained in the same form as the Ne\'{e}l ordered phase with the
replacement of the corresponding structure factors along with $\zeta\leftrightarrow 2\eta$. The explicit form for the CAF phase is given by
\begin{eqnarray}
\mathcal{M}^{\mathrm{CAF}}_{\kk,\qq}&=&2J_xS(1+2\eta)\{[1+\gamma^\prime_{1\qq}+\frac{\zeta}{1+2\eta}\nonumber\\
&&\times(\gamma^\prime_{2\kk}+\gamma^\prime_{2\kk+\qq}-\gamma^\prime_{2\qq}-1)](u^\prime_{\kk+\qq}v^\prime_{\kk}+u^\prime_{\kk}v^\prime_{\kk+\qq})\nonumber\\
&&+(\gamma^\prime_{1\kk}+\gamma^\prime_{1\kk+\qq})(u^\prime_{\kk}u^\prime_{\kk+\qq}+v^\prime_{\kk}v^\prime_{\kk+\qq})\}.
\end{eqnarray}
The frequency and momentum-dependent bimagnon scattering intensity is given by
\begin{align}
I(\qq,\omega)\propto\sum_f|\bra{i}\mathcal{O}_\qq\ket{f}|^2\delta(\omega-\omega_{fi})
=-\frac{1}{\pi}\mathrm{Im}\mathrm{G}(\qq,\omega),
\end{align}
where $\ket{i}$ and $\ket{f}$ are the initial and final states with corresponding transfered energy $\omega_{fi}$ and momentum $\qq$, respectively.
The time-ordered correlation function is given by
\begin{equation}
  \mathrm{G}(\qq,\omega)=-i\int_0^\infty\mathrm{d}t\ e^{i\omega t}\bra{i}\mathcal{T}\mathcal{O}_\qq^\dag(t)\mathcal{O}_\qq(0)\ket{i}.
\end{equation}
The momentum-dependent two-magnon Green's function is defined as
\begin{align}
\Pi(\qq,t;\kk,\kk')=-i\bra{0}\mathcal{T}\alpha_{\kk+\qq}(t)\beta_{\kk}(t)\alpha_{\kk^\prime+\qq}^\dag(0)\beta_{\kk^\prime}^\dag(0)\ket{0},
\end{align}
which can be expanded in terms of the one magnon propagators with the basic propagators
$\mathrm{G}_{\alpha\alpha}(\kk,t)=-i\bra{0}\mathcal{T}\alpha_\kk(t)\alpha_\kk^\dag(0)\ket{0}$ and
$\mathrm{G}_{\beta\beta}(\kk,t)=-i\bra{0}\mathcal{T}\beta^\dag_\kk(t)\beta_\kk(0)\ket{0}$ for the $\alpha$ and $\beta$ magnons,
where $\mathcal{T}$ is the time ordering operator and $\ket{0}$ is the ground state.
The non-interacting two-magnon propagator is denoted by $\Pi_0(\qq,\omega;\kk)$.
To include the effects of magnon-magnon interaction, we obtain a Bethe-Salpeter equation after summing the ladder diagrams exactly
\cite{PhysRevB.4.992,PhysRevB.45.7127,PhysRevB.75.214414} to obtain (see Appendix~\ref{Bethe-Salpeter} for derivation details)
\begin{align}\label{Eq:Gfull}
\mathrm{G}=\mathrm{G}_0+\hat{\mathcal{G}}\hat{\Gamma}\textbf{[}\mathbf{\hat{1}}-\mathcal{\hat{R}}\hat{\Gamma}\textbf{]}^{-1} \hat{\mathcal{G}}^{T},
\end{align}
where the individual renormalized bare correlation function is given by
\begin{align}
  \mathrm{G}_{0}(\qq,\omega)=\frac{2}{N}\sum_{\kk}\mathcal{M}_{\kk,\qq}^2\Pi_0(\qq,\omega;\kk).
\end{align}
In Eq.~(\ref{Eq:Gfull}) the matrix $\mathcal{\hat{G}}$ has dimensions of $1\times N_c$, $\hat{\mathcal{R}}(\qq,\omega)$ and the unit matrix $\hat{\mathbf{1}}$
have $N_c\times N_c$ dimensions with the matrix elements defined as
\begin{eqnarray}
\mathcal{\hat{G}}_\mathrm{m}(\qq,\omega)&=&\frac{2}{N}\sum_\kk\mathcal{M}_{\kk,\qq}v_\mathrm{m}(\kk)\Pi_0(\qq,\omega;\kk),\\
\mathcal{\hat{R}}_\mathrm{mn}(\qq,\omega)&=&\frac{2}{N}\sum_\kk v_{\mathrm{m}}(\kk)v_{\mathrm{n}}(\kk)\Pi_0(\qq,\omega;\kk).
\end{eqnarray}

\section{RIXS intensity spectrum}\label{Sec:intensity}

\begin{figure}
\centering\includegraphics[scale=0.5]{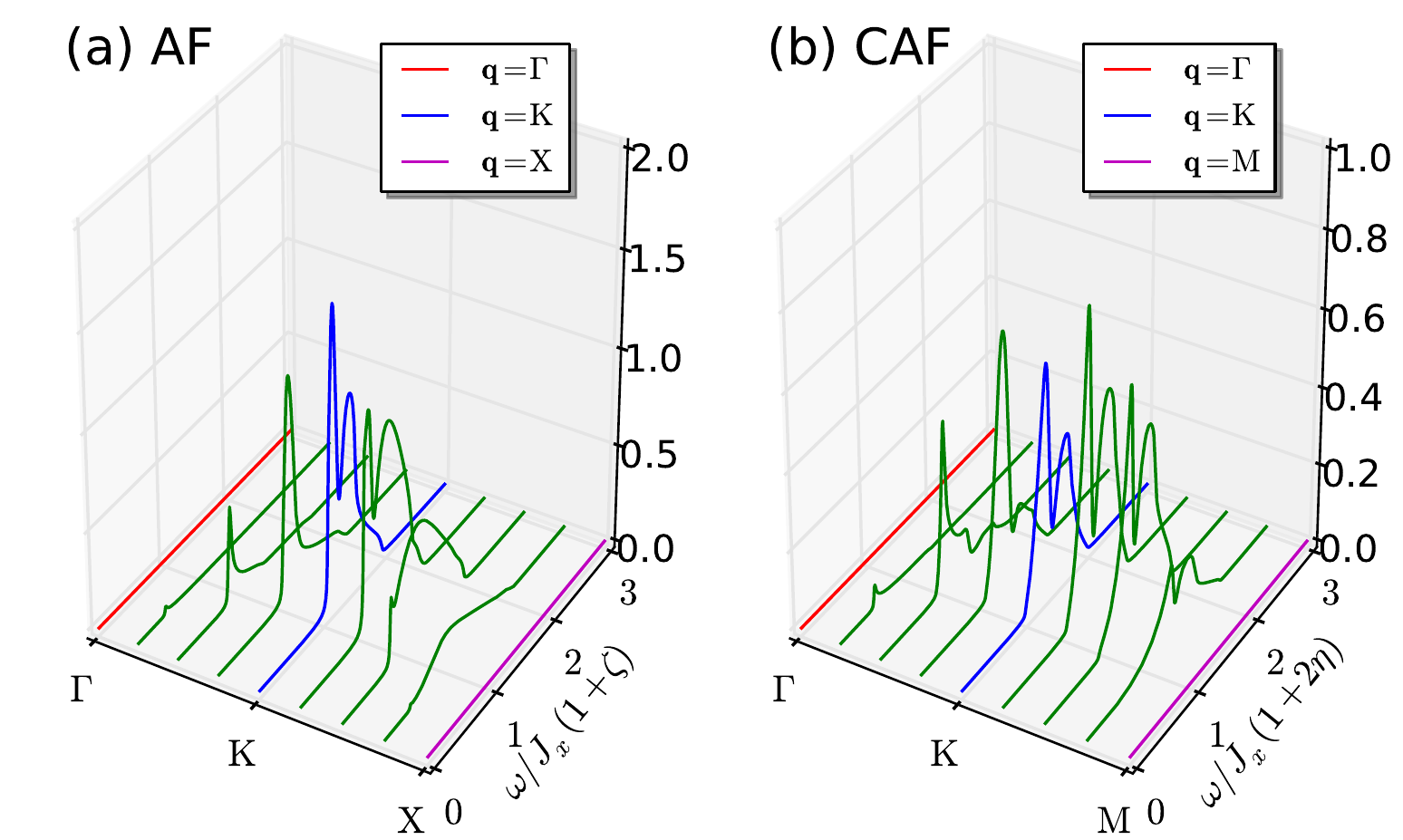}
\caption{(Color online)
Momentum-dependent bimagnon RIXS intensity of the $J_x-J_y-J_2$ model in the full spatially frustrated version
(a) AF phase ($\zeta=0.5, \eta=0.1$) and (b) CAF phase ($\zeta=0.3, \eta=0.6$) for $S=\frac{1}{2}$.
Note that the intensity vanishes at the $\Gamma$ point as well as at the antiferromagnetic wave vector
($X$ for AF phase and $M$ for CAF phase).
The robust two-peak structure appears around $K(\frac{\pi}{2},\frac{\pi}{2})$ point
for both AF and CAF phase.
}
\label{fig:afcaffall}
\end{figure}

We compute the bimagnon RIXS intensity spectra for various scattering wave vectors $\qq$ in both the AF and the CAF ordered phase.
The choice of $(\zeta,\eta)$ parameters are guided by the magnetic phase diagram of the $J_x-J_y-J_2$ model
to ensure that quantum fluctuations have not completely destroyed the two sub-lattice magnetic order.~\cite{PhysRevB.82.144407}
The classical phase diagram is given by the relation $\zeta>2\eta$ (AF) and $\zeta<2\eta$ (CAF).

In Fig.~\ref{fig:afcaffall} we display the momentum dependence of the bimagnon RIXS intensity for the full spatially frustrated $J_x-J_y-J_2$ model
in AF phase ($\zeta=0.5, \eta=0.1$)  and CAF phase ($\zeta=0.3, \eta=0.6$) with maximal quantum fluctuation $S=\frac{1}{2}$ along the
high-symmetry paths in the BZ.
The resulting magnetic RIXS spectrum shows a dispersion, which vanishes both at $\qq=\Gamma$ and $\qq=X (M)$ for the magnetically ordered
AF (CAF) phase in agreement with previous results.\cite{EPL.80.47003,PhysRevB.75.214414,PhysRevB.77.134428}
The other signature of the RIXS spectrum generated by strong spatial anisotropy and magnetic frustration is the appearance of a
two-peak structure in a wide region inside the BZ. Note that the two-peak structure is unique since the spectrum splitting will not occur either
in the unfrustrated model~\cite{EPL.80.47003,PhysRevB.75.214414,PhysRevB.77.134428} or in the large-$S$ limit where quantum fluctuations are small.

\subsection{Features of two-peak structure}\label{SubSec:peakfeature}

\begin{figure}
\centering\includegraphics[scale=0.5]{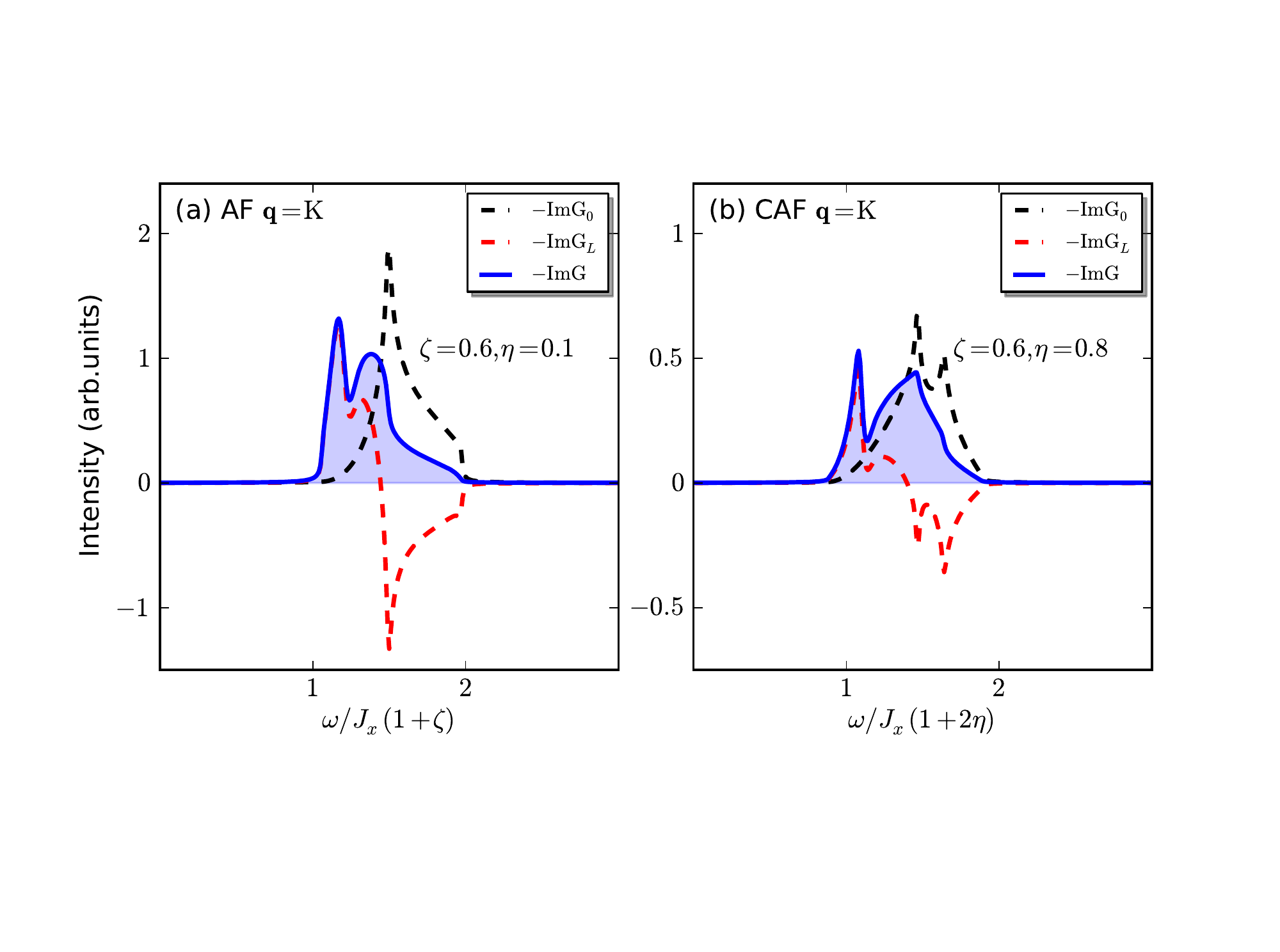}
\caption{(Color online)
Individual renormalized bare ($\mathrm{G}_{0}$, dashed black), ladder ($\mathrm{G}_{\mathrm{L}}$, dashed red),
and total interacting RIXS intensity ($\mathrm{G}$, solid blue) at $\qq=K(\frac{\pi}{2},\frac{\pi}{2})$ for $S=\half$.
Incomplete cancellation of the non-interacting contribution by the ladder interactions leads to a shoulder like feature
(AF) or to a broad peak (CAF) in the spectra.
}
\label{fig:afcafcancer}
\end{figure}

The two-peak RIXS structure in both the AF and the CAF phase has features which are worth noting. To analyze, in Fig.~\ref{fig:afcafcancer}, we show the individual renormalized bare part ($-\mathrm{ImG}_0$), ladder part ($-\mathrm{ImG_L}$), and total interacting contribution ($-\mathrm{ImG}$) to the RIXS intensity. First, the non-interacting spectra typically has Van Hove singularities shown by one or two sharp peaks. Second, the interacting RIXS spectra has a prominent peak which occurs at low energy followed by a second peak at higher energy which is either broad (CAF) or has a broad shoulder (AF). Since the non-interacting intensity generally occurs at higher energies and the interacting ladder contribution at lower energies, with a region of overlap, the ladder contributions cancel most of the non-interacting weight at high energy, but not all. This residual contribution is the reason for either the shoulder or the broad second peak in the spectra.

\subsection{Role of Anisotropy \& Frustration}\label{SubSec:Role}

\begin{figure}
\centering\includegraphics[scale=0.5]{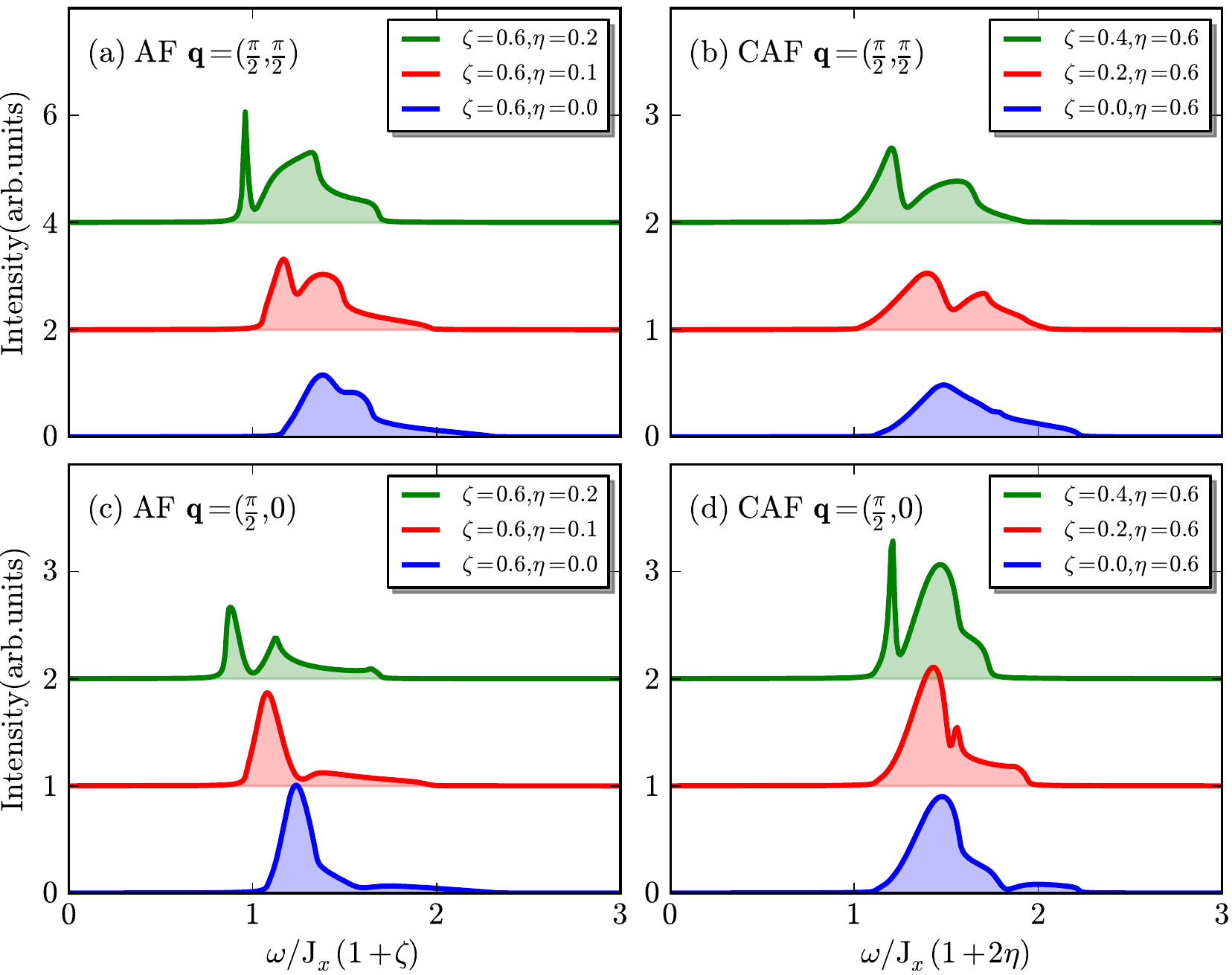}
\caption{(Color online)
Spatial anisotropy and magnetic frustration induced peak splitting of the RIXS intensity spectra obtained from the solution of the
Bethe-Salpeter equation at the designated scattering wave vector ($\qq$), spatial anisotropy parameter ($\zeta$), next-NN interaction ($\eta$),
and $S=\half$.
}
\label{fig:afcafky}
\end{figure}

The two-peak feature is a characteristic that develops at several $\qq$ vectors in the magnetic BZ.
We analyze the role of magnetic anisotropy \& frustration on the appearance of the two-peak structure of the RIXS spectra
by selecting two distinct points ($\qq=K$ and $\qq=Y$) inside the BZ where the intensity is relatively prominent.
For the AF phase, the spectrum splitting originates in the pure spatially anisotropic $J_x-J_y$ model, see Figs.~\ref{fig:afcafky}(a) and (c),
with significant peak splitting developing as the next-NN frustration increases. Another prominent feature is the asymmetry of the two-peak structure.
The low-energy branch becomes sharper with increasing frustration while the high-energy branch is broadened and spread out over a wide energy range.
Similar features are also observed in the CAF phase, as seen in Figs.~\ref{fig:afcafky}(b) and (d).
The important difference being that the single-peak structure only exists in the non-frustrated model
($\zeta\lesssim 0$). Note that in the CAF phase, the $J_y$ exchange coupling plays both the role of spatially anisotropy and magnetic frustration.
In addition, both in the AF and CAF phase, the frustrated interaction induces the spectral downshift.

\subsection{Large-$S$ analysis}\label{SubSec:Spin}

\begin{figure}
\centering\includegraphics[scale=0.5]{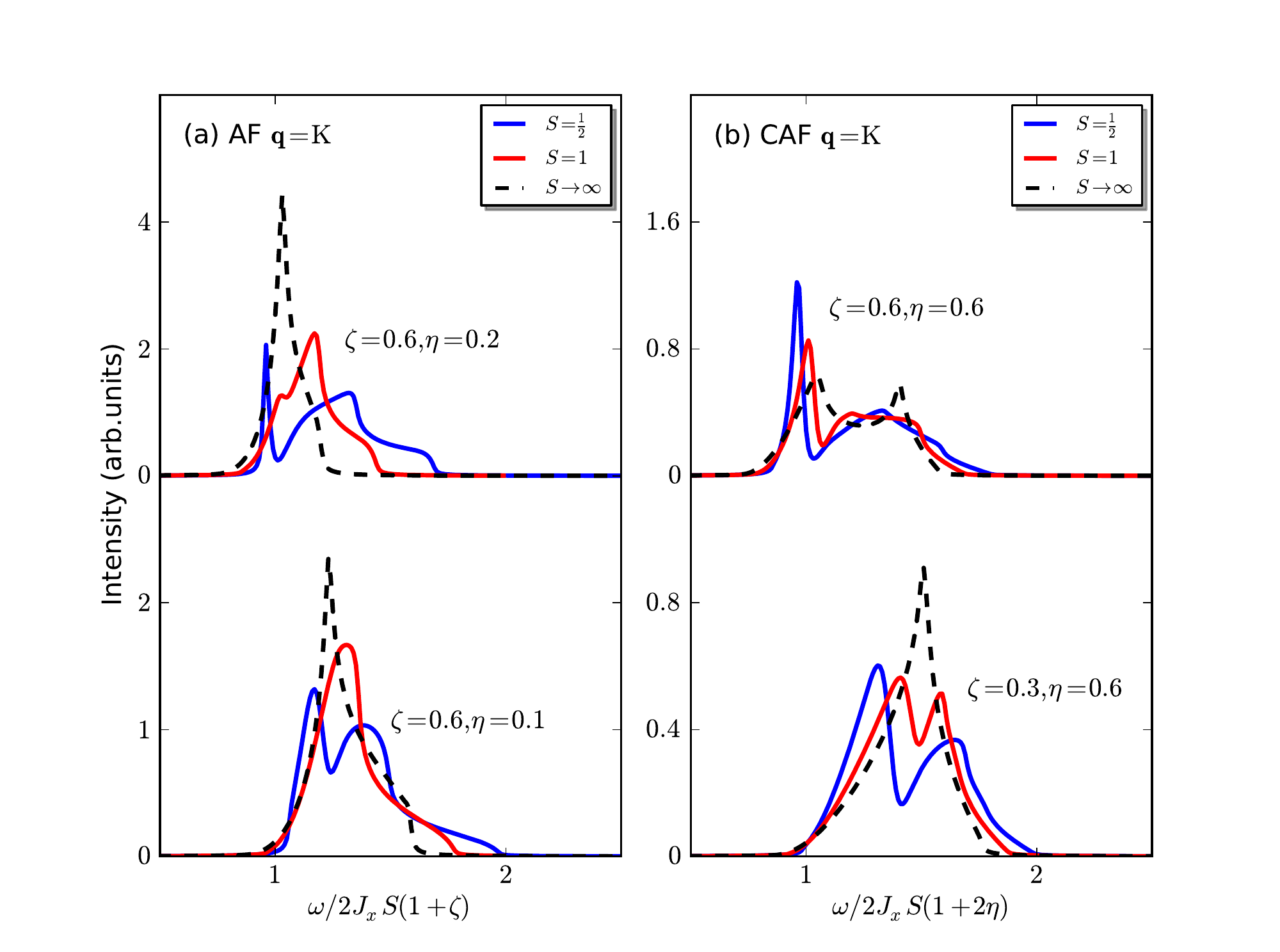}
\caption{(Color online)
Reduced quantum fluctuations from large S spin values lead to the disappearance of the two-peak structure at $K$ point for the
(a) AF and (b) CAF phase. Note that the spectrum split even in the non-interacting regime for the CAF phase with strong
enough frustration.
}
\label{fig:afcafspin}
\end{figure}

It is worth noting the effects of large-$S$ spin values on the stabilization of the two-peak structure.
In the $S\rightarrow\infty$ limit, the ladder interaction vanish and the spin wave dispersion is given by the harmonic approximation,
corresponding to two non-interacting magnons.
In Fig.~\ref{fig:afcafspin}, we show that the reduced quantum fluctuations from large S spin values can lead to the disappearance of the two-peak structure,
where we have considered the $K$ point in the BZ as an illustration. However, this feature is not generically observed for both phases.
The CAF phase can have a two-peak structure even in the non-interacting limit in the presence of strong frustration
as seen from the top panel of Fig.~\ref{fig:afcafspin}(b).

\subsection{Implications for iron pnictides}\label{SubSec:pnictides}

\begin{figure}
\centering\includegraphics[scale=0.5]{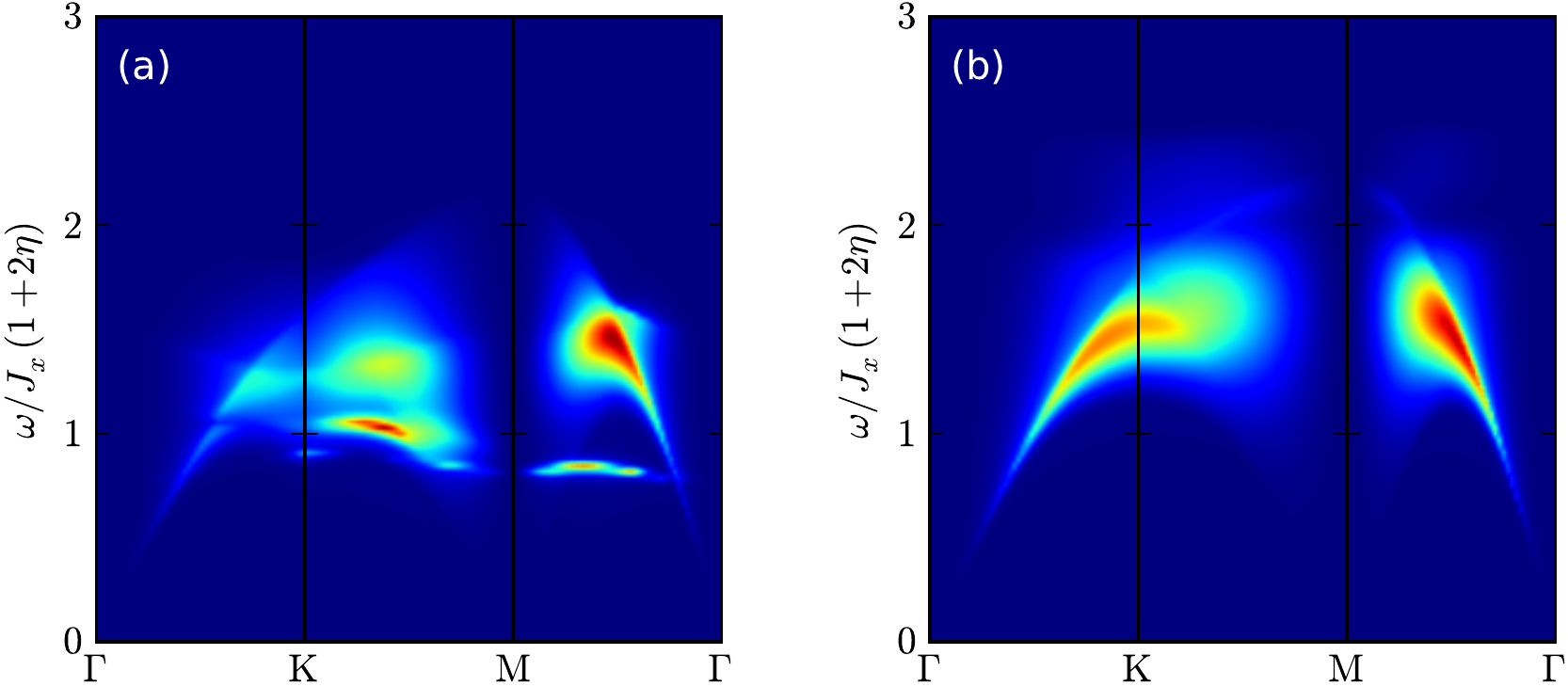}
\caption{(Color online)
Bimagnon RIXS intensity for
(a) frustrated scenario ($\zeta=0.9,\eta=1.0$) versus (b) unfrustrated scenario($\zeta=-0.1,\eta=0.4$) for iron pnictide CaFe$_2$As$_2$ ($S=\half$).
}
\label{fig:irondensity}
\end{figure}

Motivated by recent observation of antiferromagnetic correlation in the parent compound of iron pnictides,~\cite{NatPhys.5.555,NatCommun.4.1470}
we study the proposed bimagnon RIXS spectra in the CAF ordered CaFe$_2$As$_2$. To discuss the $(\pi,0)$ CAF phase, we consider
(i) a frustrated model with $\zeta=0.9,\eta=1.0$ and
(ii) the spatially unfrustrated model with $\zeta=-0.1,\eta=0.4$. Below we perform spin $S=\half$ calculations in the systems due to the relatively low
values of the staggered magnetizations measured by Neutron scattering experiments.~\cite{NatPhys.5.555}
Our results highlight the distinction between the frustrated and unfrustrated model used to describe pnictides.
In Fig.~\ref{fig:irondensity}, we show that the strongly frustrated $J_1-J_2$ model can lead to a robust spectrum splitting
along the momentum paths in the BZ, while the spectrum of the unfrustrated model typically has a single-peak structure.

\section{Comparisons with RIXS and Inelastic Neutron Scattering}\label{Sec:ins}
As mentioned in the introduction, Raman scattering is a viable complimentary probe of two magnon correlations.
Since Raman spectra is restricted to zero wave vector, it is worthwhile to inquire whether the two-peak feature survives in the Raman spectrum.
Our preliminary computations on the effects of spatial frustration and strong anisotropy indicate that there is no peak splitting,
at least in the square lattice model investigated in this paper.
A comprehensive analysis of the magnon-magnon interactions in Raman experiments for a spatially frustrated system with strong anisotropy,
especially for pnictides, and other lattice topologies is left for a future study.~\cite{unpublished}
Bimagnon spectrum can also be detected by Inelastic Neutron Scattering (INS).~\cite{PhysRevB.72.014413,PhysRevLett.105.247001}
The two-magnon spectra is linked with the longitudinal dynamical structure factor with the inelastic part of
the longitudinal structure factor directly related to the two-magnon density of states (DoS).
As point out by Lorenzana {\it et. al.},~\cite{PhysRevB.72.224511} in the case of INS the relevant scattering operator involves bosonic quasiparticles
on the same site, and the magnon-magnon interactions play a very different role than the one played in optical scattering.
A perturbation calculation can be performed in an expansion in $1/S$ and it has been shown that magnon-magnon interactions does not change
the line shape substantially in a square lattice antiferromagnet.~\cite{PhysRevB.48.3264}
The appearance of one or two strong singular peaks in the theoretical longitudinal INS spectra can not be attributed to the ladder interactions
but partially stem from the Van-Hove singularities in the two-magnon DoS.

\section{Concluding remarks}\label{Sec:Conclu}

\begin{figure}
\centering\includegraphics[scale=0.4]{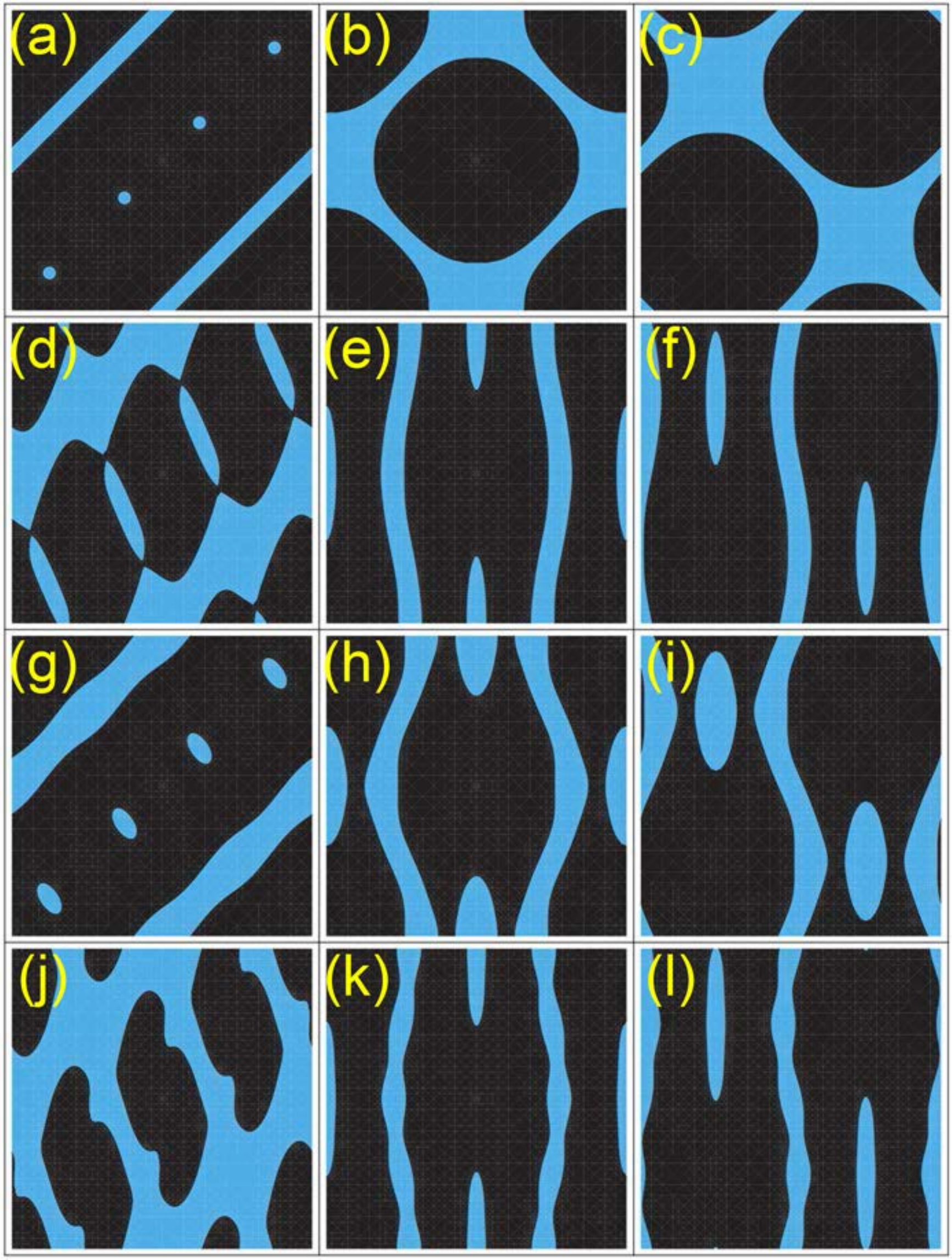}
\caption{(Color online)
AF phase single and bimagnon velocity contour plot for $S=\half$. x- and y- axis wavevector range $\in [-\pi,\pi]$.
First column: Bimagnon velocity. Second column: First magnon's velocity.
Third column: Second magnon's velocity shifted by $\qq$=($\pi\over 2$,$\pi\over 2$), i.e. $\kk_{1}$ + $\kk_{2}$ = ($\pi\over 2$,$\pi\over 2$).
Lowest velocity contours indicated by blue (grey) bands.
Relative higher velocities indicated by black regions.
Parameter choice: (a) - (c) $\zeta$=1, $\eta$=0; (d) - (f) $\zeta$=0.4, $\eta$=0.1; (g) - (h) $\zeta$=0.6, $\eta$=0.1; (j) - (l) $\zeta$=0.6, $\eta$=0.2.
}
\label{fig:velocity}
\end{figure}

We have presented a comprehensive analysis of the bimagnon {\it indirect} RIXS intensity spectra of the spatially frustrated
$J_x-J_y-J_2$ Heisenberg model in the presence of strong anisotropy and magnetic frustration.
Our treatment includes the contribution to magnon-magnon interactions at the $1/S$ order.
Similar to other spectroscopic techniques,~\cite{PhysRevB.75.024417,PhysRevB.75.020403,PhysRevB.77.174412,PhysRevB.81.085124}
the final RIXS spectra is a result of the complex interplay of ladder interaction vertex effects
and the RIXS bimagnon matrix element affected by spatial anisotropy and magnetic frustration.

Our result is significantly different from what is presently known in the RIXS community exploring quantum magnets.
For example, Forte {\it et. al.}~\cite{PhysRevB.77.134428} have carried out a linear spin wave RIXS spectra study of a far neighbor interaction
(no spatial anisotropy) Heisenberg model.
Nagao and Igarashi~\cite{PhysRevB.75.214414} go beyond the linear spin wave approach, but only to study the nearest neighbor AF model.
As our results show this is not adequate.
Experimentalists dealing with real materials need adequate guidance on materials which have strong interaction,
as highlighted by the presence of the predicted two-peak structure.

Further intuition on the appearance of the two-peak feature can be developed by tracking the effects of spatial anisotropy and magnetic
frustration on the bare \emph{bimagnon velocity}; see Fig.~\ref{fig:velocity} first column.
The individual magnon velocities are shown in the second and the third column respectively.
For the isotropic model, narrow bands and tiny pockets of low bimagnon velocities indicated by the blue (grey) spots develop as shown
in Fig.~\ref{fig:velocity}(a);
we have a single peak in this case. However, with the inclusion of spatial anisotropy and frustration the patches of
slow moving bimagnon velocity occupy greater regions of the phase space of the bimagnon continuum; see Figs.~\ref{fig:velocity}(d), (g), and (j).
Hence, the propagation of bimagnons is sensitive to the details of short range exchange couplings.
In these cases a two-peak structure appears.

A plausible explanation for the bimagnon velocity behavior, which has a direct correlation with the appearance of the two-peak RIXS spectra,
can be obtained by applying the uncertainty principle. A bimagnon is a composite object, where two magnons are at a position separation,
with a bimagnon velocity $v$. The two magnons interact with the potential energy $\mathcal{V}$ (ladder interaction) for an interaction time $\tau$.
For low bimagnon velocity (anisotropic system with frustration)
$-$ utilizing the uncertainity principle $-$ we have $\tau \sim \hbar/\mathcal{V} \ll \tau_{c}$,
where $\tau_{c}$ is the interaction collision time scale set by the ladder scattering process.
With strong interactions $\tau$ is small, setting up a high frequency of interaction. The bimagnons can participate in multiple ladder scattering events,
resulting in a split peak structure. In the absence of anisotropy or frustration, the bimagnon velocity is high as seen from Fig.~\ref{fig:velocity}.
Again, using uncertainty principle arguments we find $\tau \sim \hbar/\mathcal{V} \gg \tau_{c}$. In this scenario, the collision time is large.
There are less number of ladder scattering events (weak interaction) and the bimagnons do not lead to peak destabilization;
we find a single peak structure. Note, the bimagnon velocity trend as a signature for the appearance of a split peak is a necessary condition,
but not a sufficient one.
The trend in the magnon velocity holds for other parameter choices in the AF phase and to a certain extent in the CAF phase.
Finally, we hope that our predicted two-peak feature will encourage experimentalists to study the appearance of this fine structure detail in the
RIXS spectrum to gain insight into our understanding of correlation effects in quantum matter such as pnictides.

\begin{acknowledgments}
T.D. acknowledges invitation, hospitality, and kind support from Sun Yat-Sen University, Cottrell Research Corporation Grant,
and Georgia Regents University College of Science and Mathematics.
C.L. and D.X.Y acknowledge support from National Basic Research Program of China (2012CB821400), NSFC-11074310, NSFC-11275279,
Specialized Research Fund for the Doctoral Program of Higher Education (20110171110026), and NCET-11-0547.
D.X.Y also acknowledges discussions with Hong Ding.
\end{acknowledgments}

\appendix

\section{Separated forms of quartic ladder interaction vertex}\label{Ladder Vertex}
The channels $v_\mathrm{n}(\kk)$ and $v^\prime_\mathrm{n}(\kk)$ for the AF and CAF phase are defined in Table \ref{Table:Channel}.
\begin{table}
\caption{\label{Table:Channel}
Definition of the channels $v_\mathrm{n}(\kk)$ and $v^\prime_\mathrm{n}(\kk)$
}
\begin{ruledtabular}
\begin{tabular}{cll}
$n$ &  $v_{n}(\kk)$  &$v^\prime_{n}(\kk)$\\ \hline
 1 & $u_{\kk+\qq} u_{\kk} \cos k_x$         & $u_{\kk+\qq}^\prime u_{\kk}^\prime \cos k_x$\\
 2 & $u_{\kk+\qq} u_{\kk} \sin k_x$         & $u_{\kk+\qq}^\prime u_{\kk}^\prime \sin k_x$\\
 3 & $u_{\kk+\qq} u_{\kk} \cos k_y$         & $u_{\kk+\qq}^\prime u_{\kk}^\prime \cos k_x\cos k_y$\\
 4 & $u_{\kk+\qq} u_{\kk} \sin k_y$         & $u_{\kk+\qq}^\prime u_{\kk}^\prime \sin k_x\cos k_y$\\
 5 & $u_{\kk+\qq} v_{\kk}$                  & $u_{\kk+\qq}^\prime u_{\kk}^\prime \cos k_x\sin k_y$ \\
 6 & $v_{\kk+\qq} u_{\kk}$                  & $u_{\kk+\qq}^\prime u_{\kk}^\prime \sin k_x\sin k_y$\\
 7 & $v_{\kk+\qq} v_{\kk} \cos k_x$         & $u_{\kk+\qq}^\prime v_{\kk}^\prime$ \\
 8 & $v_{\kk+\qq} v_{\kk} \sin k_x$         & $v_{\kk+\qq}^\prime u_{\kk}^\prime$\\
 9 & $v_{\kk+\qq} v_{\kk} \cos k_y$         & $v_{\kk+\qq}^\prime v_{\kk}^\prime \cos k_x$\\
10 & $v_{\kk+\qq} v_{\kk} \sin k_y$         & $v_{\kk+\qq}^\prime v_{\kk}^\prime \sin k_x$\\
11 & $u_{\kk+\qq} v_{\kk} \cos k_x\cos k_y$ & $v_{\kk+\qq}^\prime v_{\kk}^\prime \cos k_x\cos k_y$\\
12 & $u_{\kk+\qq} v_{\kk} \sin k_x\cos k_y$ & $v_{\kk+\qq}^\prime v_{\kk}^\prime \sin k_x\cos k_y$\\
13 & $u_{\kk+\qq} v_{\kk} \cos k_x\sin k_y$ & $v_{\kk+\qq}^\prime v_{\kk}^\prime \cos k_x\sin k_y$\\
14 & $u_{\kk+\qq} v_{\kk} \sin k_x\sin k_y$ & $v_{\kk+\qq}^\prime v_{\kk}^\prime \sin k_x\sin k_y$\\
15 & $v_{\kk+\qq} u_{\kk} \cos k_x\cos k_y$ & $u_{\kk+\qq}^\prime v_{\kk}^\prime \cos k_y$ \\
16 & $v_{\kk+\qq} u_{\kk} \sin k_x\cos k_y$ & $u_{\kk+\qq}^\prime v_{\kk}^\prime \sin k_y$\\
17 & $v_{\kk+\qq} u_{\kk} \cos k_x\sin k_y$ & $v_{\kk+\qq}^\prime u_{\kk}^\prime \cos k_y$\\
18 & $v_{\kk+\qq} u_{\kk} \sin k_x\sin k_y$ & $v_{\kk+\qq}^\prime u_{\kk}^\prime \sin k_y$\\
\end{tabular}
\end{ruledtabular}
\end{table}
The matrix elements of $\hat{\Gamma}$ and $\hat{\Gamma}^\prime$ in units of $J_x$ for the AF and CAF phase are given by
\begin{widetext}
\begin{equation}
\hat{\Gamma}=
\left(
\begin{array}{cccccccccccccccccccc}
-2 & 0  & 0  & 0  & -1 & -\gamma_x^c &  0  &  0  &  0  &  0  &  0  &  0  &  0  &  0  &  0  &  0  &  0  &  0    \\
   & -2 & 0  & 0  &  0 & \gamma_x^s &  0   & 0  &  0 &  0 &  0 & 0  &  0 & 0  & 0  & 0  &  0 & 0  \\
   &  & -2\zeta & 0  & -\xi & -\zeta\gamma_y^c &  0 &  0 & 0  & 0  & 0  & 0  &  0 &  0 & 0  & 0  & 0  & 0  \\
   &  &  & 2\zeta & 0  & \zeta\gamma_y^s   &  0 &  0 & 0  &  0 & 0  &  0 &  0 & 0  & 0  &  0 & 0  & 0  \\
  &  &   &   & \theta & \phi &
-\gamma_x^c & \gamma_x^s & \zeta\gamma_y^s & \zeta\gamma_y^s &
\rho & \mu & \nu  & \chi  & 0  & 0  & 0  & 0  \\
  &  &  &  &  & \theta & -1 & 0  & -\zeta &  0 &  0 & 0  &  0 & 0  &
\rho & \mu & \nu  & \chi  \\
   &   &   &   &   &   & -2 &  0 & 0  &  0 & 0  & 0  & 0  &  0 & 0  & 0  & 0  & 0  \\
   &   &   &   &   &   &   & -2  & 0  & 0  & 0  & 0  &  0 & 0  & 0  & 0  & 0  & 0  \\
   &   &   &   &   &   &   &   & -2\zeta  & 0  & 0  & 0  & 0  &  0 & 0  & 0  & 0  & 0  \\
   &   &   &   &   &   &   &   &   & -2\zeta  & 0  &  0 & 0  &  0  & 0  &  0 & 0  & 0  \\
   &   &   &   &   &   &   &   &   &   & 4\eta & 0  & 0  & 0  & 0  & 0  & 0  & 0  \\
   &   &   &   &   &   &   &   &   &   &   & 4\eta &  0 &  0 & 0  & 0  & 0  & 0  \\
   &   &   &   &   &   &   &   &   &   &   &   & 4\eta & 0  & 0  & 0  & 0  & 0  \\
   &   &   &   &   &   &   &   &   &   &   &   &   & 4\eta & 0  & 0  & 0  & 0  \\
   &   &   &   &   &   &   &   &   &   &   &   &   &   & 4\eta &  0 & 0  & 0  \\
   &   &   &   &   &   &   &   &   &   &   &   &   &   &   & 4\eta & 0  & 0  \\
   &   &   &   &   &   &   &   &   &   &   &   &   &   &   &   & 4\eta & 0  \\
   &   &   &   &   &   &   &   &   &   &   &   &   &   &   &   &   & 4\eta \\
\end{array}
\right).
\end{equation}
\end{widetext}
In the above we have introduced the following notations
\begin{eqnarray}
 \gamma^c_{\delta}&=&\cos(\qq\cdot\bdelta),\ \gamma^s_{\delta}=\sin(\qq\cdot\bdelta),\\
 \theta&=&4\eta\gamma_{2\qq},\ \phi=-2(1+\zeta)\gamma_{1\qq},\\
 \rho&=&-2\eta+\lambda,\\
 \lambda&=&-2\eta\gamma_x^c\gamma_y^c,\ \mu=2\eta\gamma_x^s\gamma_y^c,\\
 \nu&=&2\eta\gamma_x^c\gamma_y^s,\ \chi=-2\eta\gamma_x^s\gamma_y^s.
\end{eqnarray}

\begin{widetext}
\begin{equation}
\hat{\Gamma^\prime}=
\left(
\begin{array}{cccccccccccccccccccc}
-2 & 0  &  0  & 0  & 0 &  0 &  -1  &  -\gamma_x^c  &  0  &  0  &  0  &  0  &  0  &  0  &  0  &  0  &  0  &  0    \\
   & -2 &  0  & 0  &  0 & 0 &  0   &   \gamma_x^s  &  0  &  0  &  0  & 0   &  0  & 0   & 0   & 0   &  0  & 0  \\
   &   &-4\eta& 0  & 0 & 0 & -2\eta &  \lambda & 0  & 0  & 0  & 0  &  0 &  0 & 0  & 0  & 0  & 0  \\
   &   &   & -4\eta & 0  & 0   &  0 &  \mu & 0  &  0 & 0  &  0 &  0 & 0  & 0  &  0 & 0  & 0  \\
   &   &   &   & -4\eta & \phi^\prime & 0 & \nu & 0 & 0 & 0 & 0 & 0  & 0  & 0  & 0  & 0  & 0  \\
  &  &  &  &  & -4\eta & 0 & \chi  & 0 &  0 &  0 & 0  &  0 & 0  & 0 & 0& 0 & 0\\
   &   &   &   &   &   & \theta^\prime &  \phi^\prime & -\gamma_x^c  &  \gamma_x^s & \lambda  & \mu  & \nu  &  \chi & \rho^\prime  & \zeta\gamma_y^s  & 0  & 0  \\
   &   &   &   &   &   & & \theta^\prime  & -1  & 0  & -2\eta  & 0  &  0 & 0  & 0  & 0  & \rho^\prime  & \zeta\gamma_y^s  \\
   &   &   &   &   &   &   &   & -2  & 0  & 0  & 0  & 0  &  0 & 0  & 0  & 0  & 0  \\
   &   &   &   &   &   &   &   &   & -2  & 0  &  0 & 0  &  0  & 0  &  0 & 0  & 0  \\
   &   &   &   &   &   &   &   &   &   & -4\eta & 0  & 0  & 0  & 0  & 0  & 0  & 0  \\
   &   &   &   &   &   &   &   &   &   &   & -4\eta &  0 &  0 & 0  & 0  & 0  & 0  \\
   &   &   &   &   &   &   &   &   &   &   &   & -4\eta & 0  & 0  & 0  & 0  & 0  \\
   &   &   &   &   &   &   &   &   &   &   &   &   & -4\eta & 0  & 0  & 0  & 0  \\
   &   &   &   &   &   &   &   &   &   &   &   &   &   & 2\zeta &  0 & 0  & 0  \\
   &   &   &   &   &   &   &   &   &   &   &   &   &   &   & 2\zeta & 0  & 0  \\
   &   &   &   &   &   &   &   &   &   &   &   &   &   &   &   & 2\zeta & 0  \\
   &   &   &   &   &   &   &   &   &   &   &   &   &   &   &   &   & 2\zeta \\
\end{array}
\right).
\end{equation}
\end{widetext}
In the above we have introduced the following notations
\begin{eqnarray}
 \theta^\prime&=&2\zeta\gamma^\prime_{2\qq},\ \phi^\prime=-2(1+2\eta)\gamma^\prime_{1\qq},\\
 \rho&=&-\zeta(1+\gamma_y^c).
\end{eqnarray}
Only the upper right part of the matrix is shown since the Hamiltonian is Hermitian.

\section{Solution of the ladder approximation Bethe-Salpeter equation}\label{Bethe-Salpeter}
The two-magnon Green's function can be expanded in terms of the one-magnon propagators
\begin{eqnarray}\label{Eq:Pi}
\Pi(\qq,\omega;\kk,\kk')&=&i\int\frac{\mathrm{d}\omega'}{2\pi}\mathrm{G}_{\alpha\alpha}(\kk+\qq,\omega+\omega')\nonumber\\
&&\times\mathrm{G}_{\beta\beta}(\kk,\omega')\Gamma_{\kk\kk'}(\omega,\omega'),
\end{eqnarray}
with the basic propagators up to $1/S$ order
\begin{eqnarray}
 \mathrm{G}^{-1}_{\alpha\alpha}(\kk,\omega)&=&\omega-\omega_\kk+i0^+,\\
 \mathrm{G}^{-1}_{\beta\beta}(\kk,\omega)&=&-\omega-\omega_\kk+i0^+.
\end{eqnarray}
The vertex function $\Gamma_{\kk\kk'}(\omega,\omega')$ satisfies the Bethe-Salpeter equation
\begin{eqnarray}\label{Eq:Gamma}
\Gamma_{\kk\kk'}(\omega,\omega')&=&\delta_{\kk\kk'}+i\sum_{\kk_1}\int\frac{\mathrm{d}\omega_1}{2\pi}\mathcal{V}_{\kk\kk_1}
\mathrm{G}_{\alpha\alpha}(\kk_1+\qq,\omega+\omega_1)\nonumber\\
&&\times\mathrm{G}_{\beta\beta}(\kk_1,\omega_1)\Gamma_{\kk_1\kk'}(\omega,\omega_1).
\end{eqnarray}
The unperturbed two-magnon propagator is defined as
\begin{eqnarray}\label{Eq:Pi0}
\Pi_0(\qq,\omega;\kk)&=&i\int\frac{\mathrm{d}\omega'}{2\pi}\mathrm{G}_{\alpha\alpha}(\kk+\qq,\omega+\omega')
\mathrm{G}_{\beta\beta}(\kk,\omega')\nonumber\\
&=&[\omega-\omega_{\kk+\qq}-\omega_\kk+i0^+]^{-1}.
\end{eqnarray}
It is much more convenient to directly compute the RIXS correlation function
\begin{eqnarray}\label{Eq:BS1}
\mathrm{G}(\qq,\omega)&=&\frac{2}{N}\sum_{\kk\kk'}\mathcal{M}_{\kk,\qq}\mathcal{M}_{\kk',\qq}\Pi(\qq,\omega;\kk,\kk')\nonumber\\
&=&i\frac{2}{N}\int\frac{\mathrm{d}\omega^\prime}{2\pi}\sum_\kk\mathcal{M}_{\kk,\qq}\mathrm{G}_{\alpha\alpha}(\kk+\qq,\omega+\omega')\nonumber\\
&&\times\mathrm{G}_{\beta\beta}(\kk,\omega')\Gamma_{\kk}(\omega,\omega'),
\end{eqnarray}
where we have introduced a new vertex function
\begin{equation}
\Gamma_{\kk}(\omega,\omega')=\sum\limits_{\kk'}\mathcal{M}_{\kk',\qq}\Gamma_{\kk\kk'}(\omega,\omega'),
\end{equation}
which also satisfies the following Bethe-Salpeter equation
\begin{eqnarray}\label{Eq:BS2}
\Gamma_{\kk}(\omega,\omega')&=&\mathcal{M}_{\kk,\qq}+i\sum_{\kk_1}\int\frac{\mathrm{d}\omega_1}{2\pi}\mathcal{V}_{\kk\kk_1}
\mathrm{G}_{\alpha\alpha}(\kk_1+\qq,\omega+\omega_1)\nonumber\\
&&\times \mathrm{G}_{\beta\beta}(\kk_1,\omega_1)\Gamma_{\kk_1}(\omega,\omega_1).
\end{eqnarray}
The lowest order irreducible interaction vertex in separated forms reads as
\begin{equation}
\mathcal{V}_{\kk\kk_1}=\frac{2}{N}\sum_{\mathrm{m,n}=1}^{\mathrm{N_c}} v_{\mathrm{m}}(\kk) \Gamma_{\mathrm{m n}} v_{\mathrm{n}}(\kk_1).
\end{equation}
Substituting Eq.~(\ref{Eq:BS2}) into Eq.~(\ref{Eq:BS1}), we obtain
\begin{equation}
 \mathrm{G}(\qq,\omega)=\mathrm{G}_0(\qq,\omega)+\mathcal{A}(\qq,\omega),
\end{equation}
with the non-interacting correlation function
\begin{equation}
  \mathrm{G}_{0}(\qq,\omega)=\frac{2}{N}\sum_{\kk}\mathcal{M}_{\kk,\qq}^2\Pi_0(\qq,\omega;\kk),
\end{equation}
and
\begin{widetext}
\begin{eqnarray}\label{Eq:Aomega}
\mathcal{A}(\qq,\omega)&=&\sum_{\mathrm{m,n}=1}^{\mathrm{N_c}}i\frac{2}{N}\int\frac{\mathrm{d}\omega^\prime}{2\pi}
\sum_\kk\mathcal{M}_{\kk,\qq}\mathrm{G}_{\alpha\alpha}^{(0)}(\kk+\qq,\omega+\omega')
\mathrm{G}_{\beta\beta}^{(0)}(\kk,\omega')v_\mathrm{m}(\kk)\Gamma_{\mathrm{mn}} \nonumber\\
&&\times i\frac{2}{N}\sum_{\kk_1}\int\frac{\mathrm{d}\omega_1}{2\pi}v_\mathrm{n}(\kk_1)
\mathrm{G}_{\alpha\alpha}^{(0)}(\kk_1+\qq,\omega+\omega_1)\mathrm{G}_{\beta\beta}^{(0)}(\kk_1,\omega_1)\Gamma_{\kk_1}(\omega,\omega_1)\nonumber\\
&=&\sum_{\mathrm{m,n}=1}^{\mathrm{N_c}}\mathcal{G}_{\mathrm{m}}(\qq,\omega)\Gamma_{\mathrm{mn}}\mathcal{B}_\mathrm{n}(\qq,\omega)\nonumber\\
&=&\mathcal{\hat{G}}\hat{\Gamma}\mathcal{\hat{B}}.
\end{eqnarray}
\end{widetext}
The new functions $\mathcal{\hat{G}}_\mathrm{m}(\qq,\omega)$ and $\mathcal{\hat{B}}_\mathrm{n}(\qq,\omega)$ are defined by
\begin{eqnarray}
\mathcal{\hat{G}}_\mathrm{m}(\qq,\omega)&=&\frac{2}{N}\sum_\kk\mathcal{M}_{\kk,\qq}v_\mathrm{m}(\kk)\Pi_0(\qq,\omega;\kk),\\
\mathcal{\hat{B}}_\mathrm{n}(\qq,\omega)&=&i\frac{2}{N}\sum_{\kk_1}\int\frac{\mathrm{d}\omega_1}{2\pi}v_\mathrm{n}(\kk_1)
\mathrm{G}_{\alpha\alpha}^{(0)}(\kk_1+\qq,\omega+\omega_1)\nonumber\\
&&\times\mathrm{G}_{\beta\beta}^{(0)}(\kk_1,\omega_1)\Gamma_{\kk_1}(\omega,\omega_1).
\end{eqnarray}
Recalling our basic Eq.~(\ref{Eq:BS2}) we obtain
\begin{equation}
\mathcal{\hat{B}}=\mathcal{\hat{G}}^{T}+\mathcal{\hat{R}}\hat{\Gamma}\mathcal{\hat{B}},
\end{equation}
where
\begin{equation}
\mathcal{\hat{R}}_{\mathrm{mn}}(\qq,\omega)=\frac{2}{N}\sum_\kk v_{\mathrm{m}}(\kk)v_{\mathrm{n}}(\kk)\Pi_0(\qq,\omega;\kk).
\end{equation}
We can now solve for $\mathrm{G}(\qq,\omega)$
\begin{equation}
\mathrm{G}=\mathrm{G}_0+\hat{\mathcal{G}}\hat{\Gamma}\textbf{[}\mathbf{\hat{1}}-\mathcal{\hat{R}}\hat{\Gamma}\textbf{]}^{-1} \hat{\mathcal{G}}^{T}.
\label{Eq:Gfinal}
\end{equation}
In Eq.~(\ref{Eq:Gfinal}), the matrix $\mathcal{\hat{G}}(\qq,\omega)$ are in $1\times N_c$ dimensions, $\hat{\mathcal{R}}(\qq,\omega)$ and the
unit matrix $\hat{\mathbf{1}}$ are in $N_c\times N_c$ dimensions.

\bibliographystyle{apsrev4-1}
\bibliography{refrixs}

\end{document}